\documentclass{appolb}
\usepackage{epsfig}

\def\bge{\begin{equation}}
\def\ene{\end{equation}}
\def\bg{\begin{eqnarray}}
\def\en{\end{eqnarray}}

\def\ubar{{\bar{u}}}
\def\dbar{{\bar{d}}}
\def\sbar{{\bar{s}}}

\def\bge{\begin{equation}}
\def\ene{\end{equation}}
\def\bg{\begin{eqnarray}}
\def\en{\end{eqnarray}}

\def\ubar{{\bar{u}}}
\def\dbar{{\bar{d}}}
\def\sbar{{\bar{s}}}

\begin{document}
\title{$\eta - \eta'$ mixing in $\eta$-mesic nuclei
\thanks{Presented by SDB at the International Symposium on Mesic Nuclei, 
Cracow, June 16 2010.}
}
\author{Steven D. Bass 
\address{Institute for Theoretical Physics, University of Innsbruck, \\
Technikerstrasse 25, A6020 Innsbruck, Austria}
\and
Anthony W. Thomas
\address{CSSM, School of Chemistry and Physics, \\
University of Adelaide, Adelaide SA 5005, Australia}
}
\maketitle
\begin{abstract}
$\eta$ bound states in nuclei are sensitive to the flavour-singlet component
in the $\eta$. The bigger the singlet component, the more attraction
and the greater the binding.
$\eta - \eta'$ mixing plays an important role
in understanding the value of the $\eta$-nucleon scattering length
$a_{\eta N}$.
Working with the Quark Meson Coupling model, we find a factor of two 
enhancement from mixing relative to the prediction with a pure octet $\eta$.
\end{abstract}

\section{Introduction}

Measurements of the pion, kaon and eta meson masses and their interactions
in finite nuclei provide new constraints on our understanding of dynamical
symmetry breaking in low energy QCD \cite{kienle}.
The $\eta$-nucleon interaction is attractive suggesting that $\eta$-mesons 
may form strong-interaction bound-states in nuclei.
There is presently a vigorous experimental programme to search for evidence of
these bound states \cite{pawela}.
Here we explain that
for the $\eta$
the in-medium mass $m_{\eta}^*$ 
is sensitive to the flavour-singlet
component in the $\eta$, and hence 
to the non-perturbative glue associated with axial U(1) dynamics.
An important source of the in-medium mass modification comes
from light-quarks
coupling to the scalar $\sigma$ mean-field in the nucleus 
\cite{finite0,etaqmc}.
Increasing the flavour-singlet component in the $\eta$
at the expense of the octet component gives more attraction,
more binding and a larger value of the $\eta$-nucleon
scattering length, $a_{\eta N}$ \cite{bt05}.
Since the mass shift is approximately proportional to the $\eta$--nucleon 
scattering length, it follows that that the physical value of $a_{\eta N}$ 
should be larger than if the $\eta$ were a pure octet state.

\section{QCD considerations}

Spontaneous chiral symmetry breaking suggests an octet of 
would-be Goldstone bosons:
the octet associated with chiral $SU(3)_L \otimes SU(3)_R$
plus a singlet boson associated with axial U(1)
--- each with mass squared $m^2_{\rm Goldstone} \sim m_q$.
The physical $\eta$ and $\eta'$ masses
are
about 300-400 MeV too big to fit in this picture.
One needs extra mass in the singlet channel
associated with
non-perturbative topological gluon configurations and
the QCD axial anomaly; 
-- for reviews and related phenomenology see Refs.\cite{cracow,uppsala,shore}.
\footnote
{
The QCD axial anomaly also features in discussion of the proton spin puzzle 
\cite{spin}.
}
The strange quark mass induces considerable $\eta$-$\eta'$ mixing.
For free mesons
the $\eta - \eta'$ mass matrix (at leading order in the chiral
expansion) is
\begin{equation}
M^2 =
\left(\begin{array}{cc}
{4 \over 3} m_{\rm K}^2 - {1 \over 3} m_{\pi}^2  &
- {2 \over 3} \sqrt{2} (m_{\rm K}^2 - m_{\pi}^2) \\
\\
- {2 \over 3} \sqrt{2} (m_{\rm K}^2 - m_{\pi}^2) &
[ {2 \over 3} m_{\rm K}^2 + {1 \over 3} m_{\pi}^2 + {\tilde m}^2_{\eta_0} ]
\end{array}\right)
.
\label{eq10}
\end{equation}
Here ${\tilde m}^2_{\eta_0}$ is the gluonic mass term which has a
rigorous interpretation through the Witten-Veneziano mass formula
\cite{witten,vecca}
and which
is associated with non-perturbative gluon
topology, related perhaps to confinement \cite{ks} or instantons
\cite{thooft}.
The masses of the physical $\eta$ and $\eta'$ mesons are found
by diagonalizing this matrix, {\it viz.}
\begin{eqnarray}
| \eta \rangle &=&
\cos \theta \ | \eta_8 \rangle - \sin \theta \ | \eta_0 \rangle
\\ \nonumber
| \eta' \rangle &=&
\sin \theta \ | \eta_8 \rangle + \cos \theta \ | \eta_0 \rangle
\label{eq11}
\end{eqnarray}
where
\begin{equation}
\eta_0 = \frac{1}{\sqrt{3}}\; (u\ubar + d\dbar + s\sbar),\quad
\eta_8 = \frac{1}{\sqrt{6}}\; (u\ubar + d\dbar - 2 s\sbar) 
.
\label{mixing2}
\end{equation}
One obtains values for the $\eta$ and $\eta'$ masses:
\begin{eqnarray}
m^2_{\eta', \eta} 
& &= (m_{\rm K}^2 + {\tilde m}_{\eta_0}^2 /2)
\nonumber \\
& & \pm {1 \over 2}
\sqrt{(2 m_{\rm K}^2 - 2 m_{\pi}^2 - {1 \over 3} {\tilde m}_{\eta_0}^2)^2
   + {8 \over 9} {\tilde m}_{\eta_0}^4} 
.
\nonumber \\
\label{eq12}
\end{eqnarray}
The physical mass of the $\eta$ and the octet mass
$
m_{\eta_8} = \sqrt{ {4 \over 3} m_{\rm K}^2 - {1 \over 3} m_{\pi}^2 }
$
are numerically close, within a few percent.
However, to build a theory of the $\eta$ on the octet
approximation
risks losing essential physics associated with the singlet component.

Turning off the gluonic term in Eq.(4)
one finds 
the expressions
$m_{\eta'} \sim \sqrt{2 m_{\rm K}^2 - m_{\pi}^2}$
and
$m_{\eta} \sim m_{\pi}$.
That is, without extra input from glue, in the OZI limit,
the $\eta$ would be approximately an isosinglet light-quark state
(${1 \over \sqrt{2}} | {\bar u} u + {\bar d} d \rangle$)
degenerate with the pion and
the $\eta'$ would be a strange-quark state $| {\bar s} s \rangle$
--- mirroring the isoscalar vector $\omega$ and $\phi$ mesons.
Taking the value ${\tilde m}_{\eta_0}^2 = 0.73$GeV$^2$ in the
leading-order
mass formula, Eq.(\ref{eq12}),
gives agreement with the physical masses at the 10\% level.
This value is obtained by summing over the two eigenvalues
in Eq.(4):
$
m_{\eta}^2 + m_{\eta'}^2 = 2 m_K^2 + {\tilde m}_{\eta_0}^2
$
and substituting
the physical values of $m_{\eta}$, $m_{\eta'}$ and $m_K$ \cite{vecca}.
The
corresponding
$\eta - \eta'$
mixing angle $\theta \simeq - 18^\circ$
is within the range $-17^\circ$ to $-20^\circ$ obtained
from a study of various decay processes in \cite{gilman,frere}.
The key point of Eq.(4) is that mixing and gluon dynamics play a crucial
role
in both the $\eta$ and $\eta'$ masses
and
that treating the $\eta$ as an octet pure would-be Goldstone boson risks
losing essential physics.

\section{The axial anomaly and ${\tilde m}_{\eta_0}^2$}

What can QCD tell us about the behaviour of the gluonic mass contribution
in the nuclear medium ?

The physics of axial U(1) degrees of freedom is described
by the
U(1)-extended low-energy effective Lagrangian \cite{vecca}.
In its simplest form this reads
\begin{eqnarray}
{\cal L} =
{F_{\pi}^2 \over 4}
{\rm Tr}(\partial^{\mu}U \partial_{\mu}U^{\dagger})
+
{F_{\pi}^2 \over 4} {\rm Tr} M \biggl( U + U^{\dagger} \biggr)
\nonumber \\
+ {1 \over 2} i Q {\rm Tr} \biggl[ \log U - \log U^{\dagger} \biggr]
+ {3 \over {\tilde m}_{\eta_0}^2 F_{0}^2} Q^2
.
\nonumber \\
\label{eq20}
\end{eqnarray}
Here
$
U = \exp \ i \biggl(  \Phi / F_{\pi}
                  + \sqrt{2 \over 3} \eta_0 / F_0 \biggr)
$
is the unitary meson matrix
where
$\Phi = \sum \pi_a \lambda_a$
denotes the octet of would-be Goldstone bosons associated
with spontaneous chiral $SU(3)_L \otimes SU(3)_R$ breaking
and
$\eta_0$
is the singlet boson.
In Eq.(5) $Q$ denotes the topological charge density
($Q = {\alpha_s \over 4 \pi} G_{\mu \nu} {\tilde G}^{\mu \nu}$);
$M = {\rm diag} [ m_{\pi}^2, m_{\pi}^2, 2 m_K^2 - m_{\pi}^2 ]$
is the quark-mass induced meson mass matrix.
The pion decay constant $F_{\pi} = 92.4$MeV and
$F_0$ is
the flavour-singlet decay constant,
$F_0 \sim F_{\pi} \sim 100$ MeV \cite{gilman}.

The flavour-singlet potential involving $Q$ is introduced to generate 
the gluonic contribution to the $\eta$ and $\eta'$ masses and
to reproduce the anomaly in the divergence of
the gauge-invariantly renormalised flavour-singlet axial-vector
current.
The gluonic term $Q$ is treated as a background field with no kinetic 
term. It may be eliminated through its equation of motion to generate 
a gluonic mass term for the singlet boson,
{\it viz.}
\begin{equation}
{1 \over 2} i Q {\rm Tr} \biggl[ \log U - \log U^{\dagger} \biggr]
+ {3 \over {\tilde m}_{\eta_0}^2 F_{0}^2} Q^2
\
\mapsto \
- {1 \over 2} {\tilde m}_{\eta_0}^2 \eta_0^2
.
\label{eq23}
\end{equation}

The interactions of the $\eta$ and $\eta'$ with other mesons and
with nucleons can be studied by coupling the Lagrangian Eq.(5) to
other particles \cite{bass99,veccb}.
For example,
the OZI violating interaction
$\lambda Q^2 \partial_{\mu} \pi_a \partial^{\mu} \pi_a$
is needed to generate the leading (tree-level)
contribution to the decay $\eta' \rightarrow \eta \pi \pi$
\cite{veccb}.
When iterated in the Bethe-Salpeter equation for meson-meson
rescattering
this interaction yields a dynamically generated exotic state
with quantum numbers $J^{PC} = 1^{-+}$ and mass about 1400 MeV
\cite{bassmarco}.
This suggests a dynamical interpretation of the lightest-mass 
$1^{-+}$ exotic observed at BNL and CERN.

To investigate what happens to ${\tilde m}^2_{\eta_0}$ in the medium
we first couple
the $\sigma$
(correlated two-pion)
mean-field in nuclei
to the topological charge density $Q$
through adding the Lagrangian term
\begin{equation}
{\cal L}_{\sigma Q} =
Q^2 \ g_{\sigma}^Q \sigma
\label{eq27}
\end{equation}
Here
$g_{\sigma}^Q$ denotes coupling to the $\sigma$ mean field
--
that is, we
consider an in-medium renormalization of the coefficient of $Q^2$
in the effective chiral Lagrangian.
Following the treatment in Eq.(6) we eliminate
$Q$ through its equation of motion.
The gluonic mass term for the singlet boson then becomes
\begin{equation}
{\tilde m}^2_{\eta_0}
\mapsto
{\tilde m}^{*2}_{\eta_0}
=
{\tilde m}^2_{\eta_0}
\ { 1 + 2 x \over (1 + x)^2 }
\ < {\tilde m}^2_{\eta_0}
\label{eq28}
\end{equation}
where
\begin{equation}
x =
{1 \over 3} g_{\sigma}^Q \sigma \ {\tilde m}^2_{\eta_0} F_0^2.
\label{eq29}
\end{equation}
That is, {\it the gluonic mass term decreases in-medium}
independent of the sign of $g_{\sigma}^Q$ and the medium acts
to partially neutralize axial U(1) symmetry breaking by gluonic effects.

This discussion motivates the {\it existence} of 
medium modifications to ${\tilde m}^2_{\eta_0}$ in QCD.
\footnote{
In the chiral limit the singlet
analogy to the Weinberg-Tomozawa
term does not vanish because of the anomalous glue terms.
Starting from the simple Born term one finds
anomalous gluonic contributions
to the singlet-meson nucleon scattering length
proportional to ${\tilde m}^2_{\eta_0}$ and ${\tilde m}_{\eta_0}^4$
\cite{bassww}.
}
However, a rigorous calculation of $m_{\eta}^{*}$ from QCD 
is beyond present theoretical technology. 
Hence, one has to look to QCD motivated models and phenomenology for
guidance about the numerical size of the effect.
The physics described in 
Eqs.(1-4) tells us that the simple octet approximation may not suffice.

\section{The $\eta$ in nuclei}

\subsection{QCD inspired Models}

Meson masses in nuclei are determined from the scalar induced contribution 
to the meson propagator evaluated at zero three-momentum, ${\vec k} =0$, in 
the nuclear medium.
Let $k=(E,{\vec k})$ and $m$ denote the four-momentum and mass of the meson 
in free space.
Then, one solves the equation
\begin{equation}
k^2 - m^2 = {\tt Re} \ \Pi (E, {\vec k}, \rho)
\end{equation}
for ${\vec k}=0$
where $\Pi$ is the in-medium $s$-wave meson self-energy.
Contributions to the in medium mass come from coupling to the scalar 
$\sigma$ field in the nucleus in mean-field approximation,
nucleon-hole and resonance-hole excitations in the medium.
The $s$-wave self-energy can be written as \cite{ericson}
\begin{equation}
\Pi (E, {\vec k}, \rho) \bigg|_{\{{\vec k}=0\}}
=
- 4 \pi \rho \biggl( { b \over 1 + b \langle {1 \over r} \rangle } \biggr) .
\end{equation}
Here $\rho$ is the nuclear density,
$
b = a ( 1 + {m \over M} )
$
where 
$a$ is the meson-nucleon scattering length, $M$ is the nucleon mass and
$\langle {1 \over r} \rangle$ is
the inverse correlation length,
$\langle {1 \over r} \rangle \simeq m_{\pi}$ 
for nuclear matter density \cite{ericson}.
($m_{\pi}$ is the pion mass.) 
Attraction corresponds to positive values of $a$.
The denominator in Eq.(11) is the Ericson-Ericson-Lorentz-Lorenz
double scattering correction.

What should we expect for the $\eta$ and $\eta'$ ?

This physics with $\eta - \eta'$ mixing has been investigated 
by Bass and Thomas \cite{bt05}.
Phenomenology is used
to estimate the size of the effect in the $\eta$
using
the Quark Meson Coupling model (QMC) of hadron properties in the nuclear 
medium \cite{etaqmc}.
Here one uses the large $\eta$ mass 
(which in QCD is induced by mixing and the gluonic mass term)
to motivate taking an MIT Bag 
description
for the $\eta$ wavefunction, and
then coupling the light (up and down)
quark and antiquark fields in the $\eta$ to the scalar $\sigma$
field
in the nucleus working in mean-field approximation \cite{etaqmc}.
The coupling constants in the model for the coupling of light-quarks
to the $\sigma$ (and $\omega$ and $\rho$) mean-fields in the nucleus
are
adjusted to fit the saturation energy and density of
symmetric nuclear matter and the bulk symmetry energy.
The strange-quark component of the wavefunction does not couple
to the $\sigma$ field and $\eta-\eta'$ mixing is readily built into the model.

Increasing the mixing angle increases the amount of singlet 
relative to octet components in the $\eta$.
This produces greater attraction through increasing 
the amount of light-quark compared to strange-quark 
components in the $\eta$
and a reduced effective mass.
Through Eq.(11), increasing the mixing angle 
also increases 
the
$\eta$-nucleon scattering length $a_{\eta N}$.
The model results are shown in Table 1.
The values of ${\tt Re} a_{\eta}$ quoted in Table 1 are obtained
from substituting the in-medium and free masses into Eq.(11) with
the Ericson-Ericson denominator turned-off
(since we choose to work in mean-field approximation), and using 
the free
mass $m=m_{\eta}$
in the expression for $b$.
\footnote{The effect of exchanging $m$ for
$m^*$ in $b$ is a 5\% increase in the quoted scattering length.}
The QMC model makes no claim about the imaginary part of the scattering
length.
The key observation is that $\eta - \eta'$ mixing 
with the phenomenological mixing angle $-20^\circ$
leads to a factor of two increase in the mass-shift and 
in the scattering length obtained in the model
relative to the prediction for a pure octet $\eta_8$.
This result may explain why values of $a_{\eta N}$ extracted from 
phenomenological fits to experimental data where the $\eta-\eta'$ 
mixing angle is unconstrained 
give larger values than those predicted 
in theoretical models where the $\eta$ is treated as a pure octet state
-- see below.

\begin{table}[t!]
\begin{center}
\caption{
Physical masses fitted in free space, 
the bag 
masses in medium at normal nuclear-matter 
density,
$\rho_0 = 0.15$ fm$^{-3}$, 
and corresponding meson-nucleon scattering lengths
(calculated at the mean-field level 
 with the Ericson-Ericson-Lorentz-Lorenz factor switched off).
}
\label{bagparam}
\begin{tabular}[t]{c|lll}
\hline
&$m$ (MeV) 
& $m^*$ (MeV) & ${\tt Re} a$ (fm)
\\
\hline
$\eta_8$  &547.75  
& 500.0 &  0.43 \\
$\eta$ (-10$^o$)& 547.75  
& 474.7 & 0.64 \\
$\eta$ (-20$^o$)& 547.75  
& 449.3 & 0.85 \\
$\eta_0$  &      958 
& 878.6  & 0.99 \\
$\eta'$ (-10$^o$)&958 
& 899.2 & 0.74 \\
$\eta'$ (-20$^o$)&958 
& 921.3 & 0.47 \\
\hline
\end{tabular}
\end{center}
\end{table}

The density dependence of the mass-shifts in the QMC model is discussed
in Ref.\cite{etaqmc}.
Neglecting the Ericson-Ericson term, the mass-shift is approximately
linear
For densities $\rho$ between 0.5 and 1 times $\rho_0$ (nuclear
matter density) we find
\begin{equation}
m^*_{\eta} / m_{\eta} \simeq 1 - 0.17 \rho / \rho_0
\end{equation}
for the mixing angle $-20^\circ$.
The scattering lengths extracted from this analysis are density independent 
to within a few percent over the same range of densities.

Present experiments \cite{pawela} are focussed on searches 
for $\eta$-mesic Helium.
QMC model calculations for finite nuclei are reported in \cite{etaqmc}.
For an octet eta, $\eta_8$, 
one finds a binding energy of 10.7 MeV in $^6$He.
(This binding energy is expected to double with $\eta -\eta'$ mixing 
included.)
Calculations of the $\rho$-meson mass in $^3$He and $^4$He are reported
in \cite{rhoqmc}. 
One finds that the average mass for a $\rho$-meson formed in $^3$He and 
$^4$He is expected to be around 730 and 690 MeV.

\subsection{Comparison with $\eta$ phenomenology and other models}

It is interesting to compare these results with other studies and
the values of
$a_{\eta N}$ and $a_{\eta' N}$
extracted from phenomenological fits to experimental data.

The $\eta$-nucleon interaction is characterised by a strong coupling 
to the $S_{11}$(1535) nucleon resonance.
For example, eta meson production in proton nucleon collisions 
close to threshold is known 
to procede via a strong isovector exchange contribution with 
excitation of the $S_{11}(1535)$.
Recent measurements of etaprime production suggest a different
mechanism for this meson \cite{pawelcosy11}.
Different model procedures lead to different values of the 
$\eta$-nucleon
scattering length with real part between about 0.2fm and 0.9fm.

In quark models the $S_{11}$ is interpreted as a 3-quark state: $(1s)^2(1p)$.
This interpretation has support from quenched lattice calculations
\cite{lattice}
which also suggest that the $\Lambda (1405)$ resonance has a significant non
3-quark component.
In the Cloudy Bag Model the $\Lambda (1405)$ is dynamically generated in the
kaon-nucleon system \cite{CBM}.

{\it Phenomenological determinations of $a_{\eta N}$ and $a_{\eta' N}$:}
Green and Wycech \cite{wycech} have performed phenomenological
K-matrix
fits to a variety of near-threshold processes
($\pi N \rightarrow \pi N$, $\pi N \rightarrow \eta N$,
 $\gamma N \rightarrow \pi N$ and $\gamma N \rightarrow \eta N$)
to extract a value for the $\eta$-nucleon scattering.
In these fits the $S_{11}(1535)$ is introduced as an explicit
degree of freedom
-- that is, it is treated like a 3-quark state --
and the $\eta-\eta'$ mixing angle is taken as a free parameter.
The real part of $a_{\eta N}$ extracted from these fits is 0.91(6) fm
for the on-shell scattering amplitude.

From measurements of $\eta$ production in proton-proton collisions
close to threshold,
COSY-11 have extracted a scattering length
$a_{\eta N} \simeq 0.7$ + i 0.4fm
from the final state interaction (FSI)
based on the effective range approximation
 \cite{cosyeta}.
For the $\eta'$, COSY-11 have deduced a
conservative upper bound on
the $\eta'$-nucleon scattering length
$| {\tt Re} a_{\eta' N} | < 0.8$fm \cite{cosy}
with a prefered a value between 0 and 0.1 fm \cite{pawel}
obtained by comparing the FSI in $\pi^0$ and $\eta'$ production
in proton-proton collisions close to threshold.

{\it Chiral Models:}
Chiral models involve performing a coupled channels analysis of
$\eta$ production after multiple rescattering in the nucleus
which is calculated
using the Lippmann-Schwinger \cite{etaweise} or Bethe-Salpeter
\cite{etaoset} equations with potentials taken from the SU(3)
chiral Lagrangian for low-energy QCD.
In these chiral model calculations
the $\eta$ is taken as pure octet state
$(\eta = \eta_8)$ with no mixing and the singlet sector turned off.
These calculations
yield a small mass shift in nuclear matter
$
m^*_{\eta} / m_{\eta} \simeq 1 - 0.05 \rho / \rho_0
$.
The values of the $\eta$-nucleon scattering length extracted from
these chiral model calculations are
0.2 + i 0.26 fm
\cite{etaweise} and
0.26 + i 0.24 fm
\cite{etaoset}
with slightly different treatment of the intermediate state mesons.
Chiral coupled channels models with an octet $\eta = \eta_8$ agree 
with
lattice and Cloudy Bag model predictions 
for the $\Lambda (1405)$ 
and differ for the $S_{11} (1535)$, 
which is interpreted as a $K \Sigma$ 
quasi-bound state in these coupled channel calculations \cite{kaiser}.

\section{CONCLUSIONS}

$\eta - \eta'$ mixing plays a vital role in the $\eta$-nucleon and
-nucleus interactions.
The greater the flavour-singlet component in the $\eta$,
the greater the $\eta$ binding energy in nuclei through
increased attraction and the smaller the value of $m_{\eta}^*$.
Through Eq.(11), this corresponds to an increased $\eta$-nucleon
scattering length $a_{\eta N}$,
greater than the value one would expect if the $\eta$ were a pure octet state.
Measurements of $\eta$ bound-states in nuclei
are therefore a probe of singlet axial U(1)
dynamics in the $\eta$.

\vspace{1.0cm}

{\bf Acknowledgements} \\

We thank K. Tsushima for helpful communications.
SDB thanks
P. Moskal for the invitation to talk at this stimulating meeting. 
The research of
SDB is supported by the Austrian Science Fund, FWF, through grant
P20436,
while
AWT is supported by the Australian Research Council through an
Australian Laureate Fellowship and by the University of Adelaide.

\end{document}